\documentclass[acmtog, authorversion]{acmart}
\usepackage{graphicx}

\usepackage{calc}
\usepackage{enumitem}
\usepackage{booktabs}

\AtBeginDocument{%
  \providecommand\BibTeX{{%
    \normalfont B\kern-0.5em{\scshape i\kern-0.25em b}\kern-0.8em\TeX}}}

\copyrightyear{2021}
\acmYear{2021}
\setcopyright{acmlicensed}\acmConference[DIS '21]{Designing Interactive
Systems Conference 2021}{June 28-July 2, 2021}{Virtual Event, USA}
\acmBooktitle{Designing Interactive Systems Conference 2021 (DIS '21), June
28-July 2, 2021, Virtual Event, USA}
\acmPrice{15.00}
\acmDOI{10.1145/3461778.3462134}
\acmISBN{978-1-4503-8476-6/21/06}

\begin{document}

\title[Eliciting New Perspectives in RtD Studies through Annotated Portfolios]{Eliciting New Perspectives in RtD Studies through Annotated Portfolios: A Case Study of Robotic Artefacts}

\author{Marius Hoggenmueller}
\authornote{Both authors contributed equally to this research.}
\author{Wen-Ying Lee}
\authornotemark[1]
\affiliation{
  \institution{Design Lab, Sydney School of Architecture, Design and Planning, The University of Sydney}
  \country{Australia}
}
\affiliation{
  \institution{Robots in Groups Lab, Cornell University}
  \country{United States of America}
}
\email{marius.hoggenmueller@sydney.edu.au}
\email{wl593@cornell.edu}

\author{Luke Hespanhol}
\affiliation{%
  \institution{Design Lab, Sydney School of Architecture, Design and Planning, The University of Sydney}
  \country{Australia}
}
\email{luke.hespanhol@sydney.edu.au}

\author{Malte Jung}
\affiliation{
\institution{Robots in Groups Lab, Cornell University}
\country{United States of America}
}
\email{mfj28@cornell.edu}

\author{Martin Tomitsch}
\affiliation{%
  \institution{Design Lab, Sydney School of Architecture, Design and Planning, The University of Sydney}
  \country{Australia}
}
\affiliation{%
  \institution{CAFA Beijing Visual Art Innovation Institute}
  \country{China}
}
\email{martin.tomitsch@sydney.edu.au}

\renewcommand{\shortauthors}{Hoggenmueller and Lee, et al.}

\begin{abstract}
In this paper, we investigate how to elicit new perspectives in research-through-design (RtD) studies through annotated portfolios.
Situating the usage in human-robot interaction (HRI), we used two robotic artefacts as a case study: we first created our own annotated portfolio and subsequently ran online workshops during which we asked HRI experts to annotate our robotic artefacts. 
We report on the different aspects revealed about the value, use, and further improvements of the robotic artefacts through using the annotated portfolio technique ourselves versus using it with experts. 
We suggest that annotated portfolios -- when performed by external experts -- allow design researchers to obtain a form of creative and generative peer critique. 
Our paper offers methodological considerations for conducting expert annotation sessions. Further, we discuss the use of annotated portfolios to unveil designerly HRI knowledge in RtD studies.
\end{abstract}

\ccsdesc[500]{Human-centered computing~Interaction design theory, concepts and paradigms}
\ccsdesc[500]{Human-centered computing~Empirical studies in interaction design}
\keywords{annotated portfolios, research through design, intermediate-level knowledge, human-robot interaction, urban robots}

\begin{teaserfigure}
  \includegraphics[width=\textwidth]{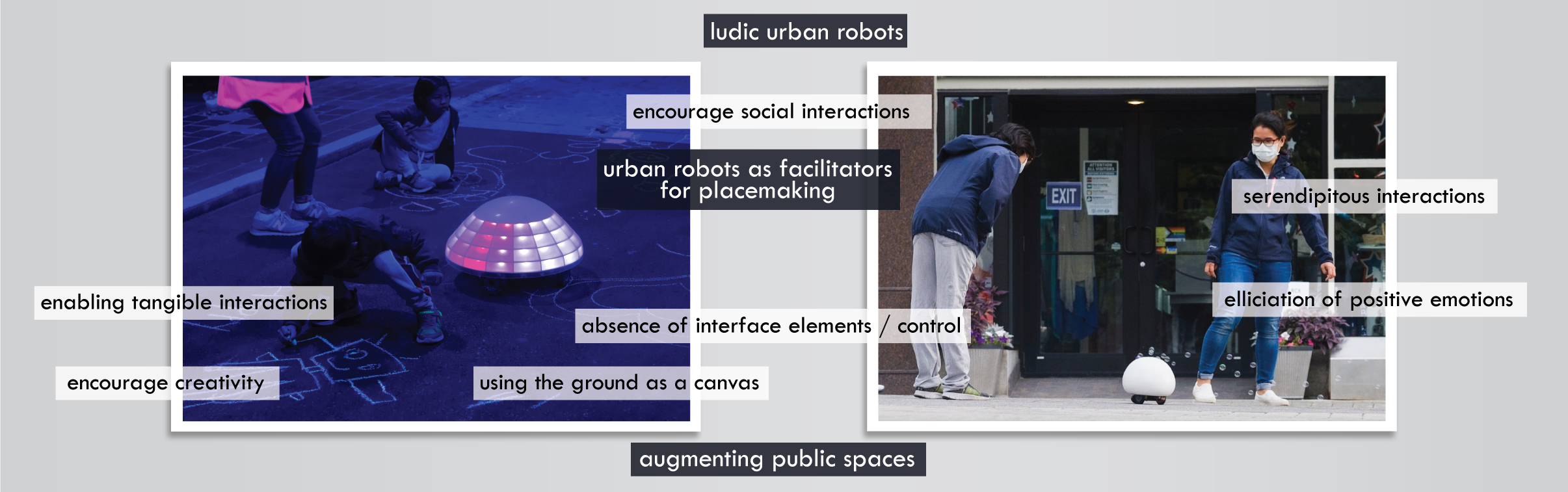}
  \caption{Woodie (left) and BubbleBot (right) with individual and shared annotations concerning interaction qualities (white) and the design domain (black).}
  \label{fig:designer_ap1}
\end{teaserfigure}

\maketitle

\section{Introduction}

Research through Design (RtD), described by Zimmerman et al. as ``an approach that employs methods and processes from design practices'' \cite{zimmerman2007research}, has been established and continued to grow in human-computer interaction (HCI). Despite receiving criticism for a lack of standardisation, rigour and an over-saturation with design artefacts \cite{Zimmerman2008, Zimmerman2010, Fallman2010}, RtD is advocated by researchers for making contributions by addressing under-constrained problems and understanding a broader design context \cite{Cullen2020}. In his essay on what to expect from RtD, Gaver urged caution on calls for convergence and standardisation, which would undermine the ability of design to create multiple possibilities \cite{gaver2012should}. Instead, he advocated for considering ``theory as annotation of realised design examples''. In this vein, researchers have discussed various intermediate-level knowledge forms, such as annotated portfolios \cite{Lowgren2013}, strong concepts \cite{Hook2012}, design patterns \cite{Duyne2002}, and bridging concepts \cite{Dalsgaard2014}. These forms of knowledge reside in between particular design instances and abstract theory. 

Over the past decade, the DIS community has significantly contributed towards the discussion of intermediate-level knowledge, leading to the acceptance of RtD approaches in the wider HCI community. However, there are subfields in HCI, as well as new emerging design domains with roots originated in engineering and technical sciences, where design research is still underrepresented. For example, recently, researchers in human-robot interaction (HRI) started to discuss the contributions that RtD work could bring to the field \cite{Luria2019, LuriaHRI21} and what design epistemology in HRI could be \cite{workshop, Lupetti2021}. Although new methods building on speculative and critical design approaches have been proposed (e.g. futuristic autobiographies \cite{Cheon2018} to elicit values and ethics around robots) and practice-based design work in robotics exists (see ``Technological Dreams Series: No.1, Robots'' by Dunne and Raby \cite{Dunne2007}), questions remain about how to generate and document knowledge from particular robotic artefacts and the underlying process of making. Building on the concept of ``designerly ways of knowing'' and thinking \cite{cross1982designerly}, Lupetti et al. \cite{Lupetti2021} contextualised intermediate-level knowledge in HRI and presented a toolbox for what they refer to as ``designerly HRI'' work. They proposed the usage of annotated portfolios, amongst others, to conceptualise knowledge from a collection of robotic artefacts, yet at the same time stressing that examples and a systematic investigation are still missing in the field.

In this paper, we continue the line of discussion and specifically look into the usage and adaptation of annotated portfolios through two of our own RtD studies on urban robots. Our paper is composed of two main parts: First, we created an annotated portfolio from a designer's perspective of our robotic artefacts to demonstrate how the approach can support the understanding of the conceptual implications of designerly HRI work. We adopted annotation strategies identified in previous work (e.g. regarding interaction qualities and domain \cite{Gaver2012, bowers2012logic}) and identified new strategies, such as zooming in on specific interface aspects as well as contrasting the designer's initial ideals with the actual outcome. Second, we conducted online workshops with invited experts\footnote{For the sake of simplicity, we will be using the term 'expert' to refer to 'external expert' throughout the rest of the paper.} from the HRI design field to create annotated portfolios on the same two robotic artefacts. The experts were external in the sense that they were not in any form involved in the RtD studies and the underlying design process itself. This represents a novel application of annotated portfolios, akin to peer-review, in the sense that the design researchers can gain new insights on their own cases through the external perspectives brought in by other experts. 

In the course of this paper, we reflect on each of the strategies from our own portfolio and examine the experts' annotations to discuss how annotated portfolios contribute to articulating the conceptual implications of robotic artefacts. 
We further discuss the different aspects revealed through using annotated portfolios with experts and present methodological considerations to deploy expert annotation sessions.
Finally, we look into the specific use of annotated portfolios in HRI context and how the approach can contribute to abstracting designerly knowledge in RtD studies.
Using two robotic artefacts as a case study, our paper exemplifies the value of designerly knowledge in HRI. Further, it adds to previous work on annotated portfolios that we hope is valuable for the broader interaction design community concerned with the articulation of knowledge gained from design research.

\section{Related Work}
Our paper builds on and contributes to previous work in HCI on intermediate-level knowledge and annotated portfolios, which we extend to the HRI context. 

\subsection{Intermediate Forms of Knowledge in Design Research}
Following the increasing work on interaction design in HCI, the knowledge-oriented discourse has started to grow in the community to better position and address design practices \cite{Hook2012}. Even though RtD has established its recognition in the field, design researchers are still facing the struggle of generalising to ``scientific theory'' \cite{Gaver2012, bowers2012logic, Lowgren2013}. To this end, intermediate-level knowledge was introduced as the middle territory between design instances and theory \cite{Hook2012}. The concept represents the new understanding towards design theory and pays attention to the various forms of knowledge being produced in design as the embodiment of "designerly" ways of knowing \cite{cross1982designerly}. 

While forms of intermediary design knowledge are proposed, the need for better ways to communicate, contest, and develop them in academia still remains \cite{Hook2015knowledge, Hook2015framing}. In this light, Bardzell et al. argued that documentation of RtD process is the key to translating design knowledge into broader academic knowledge since the knowledge is ``embodied in the object'' \cite{Bardzell2016}. They proposed viewing annotated portfolios as a ``genre of RtD aggregations-as-discourse''. Their idea resonates with L\"owgren's promotion toward annotated portfolios, mapping them out as an approach that can relate to other forms, such as \textit{patterns}, \textit{strong concepts}, and \textit{experiential qualities}, to build the stronger interlinked web of intermediate-level knowledge \cite{Lowgren2013}. 

\subsection{Annotated Portfolios}
The notion of annotated portfolios was originally introduced by Gaver and Bowers as a method to communicate design research \cite{Gaver2012}. They exemplified the approach by representing a visual collection of artefacts from their own practice, combined with brief textual annotations to outline similarities and family resemblances within the works. Their intention was to provide designers with an approach to articulate design as research, however, not relinquish the particularity of the design work, and instead esteem existing practices. In further elaborations, Bowers \cite{bowers2012logic} formulated an initial set of features of annotated portfolios, emphasising their descriptive yet generative-inspirational nature to highlight, formulate, and collate design thinking. While Bowers argues for annotations being purely indexical rather than abstractions, L\"owgren puts forward that through annotating a collection of artefacts, the portfolio naturally reaches a level of abstraction in terms of a wider applicability \cite{Lowgren2013}. Further, L\"owgren elaborates on the perspective of the designer compared to the recipient of the portfolio, arguing that even if annotations are not intended as abstractions they might be perceived and appropriated as such.

Given the initial intention was to offer an approach that is open to interpretation \cite{Gaver2012}, aligned with what has been proposed by Sengers and Gaver in their earlier work \cite{Sengers}, design researchers have adapted and expanded the usage of annotated portfolios in various ways. This includes, for example, the choice of dissemination form and representation style (e.g. purely text-based \cite{bowers2012logic}, predominantly visual \cite{jarvis2012attention}, or a combination thereof \cite{Gaver2012, Cullen2020}). Further, portfolios can be annotated to examine different foci (i.e. what the designer aims to communicate \cite{bowers2012logic}) by applying different strategies. While the majority of works followed strategies that shed light on ``interaction qualities'' and ``domain knowledge'' embodied in a collection of artefacts \cite{Cullen2020}, new approaches have also been put forward: Culén et al., for example, proposed the ``design trajectory'' and ``design ecosystem'' strategies to highlight the chronological development of artefacts and respectively how artefacts can complement each other \cite{Cullen2020}.

Annotated portfolios have also been appropriated to serve additional purposes. For instance, Hauser et al. applied annotated portfolios to demonstrate how inquiries through research products can be seen as an experimental way of doing postphenomenology in HCI, thereby tracing methodological commitments \cite{hauser2018annotated}. Others have used annotated portfolios to retrospectively reflect on considerations made during the design process \cite{jarvis2012attention, Hoang2018what}, to propose real-time annotations to document design activities \cite{rasmussen2019co}, or to get insights during design ideation sessions with prospective users \cite{hsu2018botanical}. The later raises interesting questions whether annotated portfolios -- as proposed by Gaver -- have to be performed by the originating designer or can be performed by other experts as well. Lockton et al. have briefly mentioned how annotations made by the users on the research artefact can provide qualitative feedback to the designer \cite{lockton2020sleep}. In the work of Luciani et al. on curated collections, they stressed on having curators as a source to provide meaningful annotations and reflection as a form of critique \cite{Luciani2018MachineLA}. Following up on this, Lupetti et al. also raised the question about the potential outcome and contribution through annotating other people's work in HRI \cite{Lupetti2021}.

To date, annotated portfolios have been applied to engage multiple meanings in design and document the design thinking through knowledge-abstraction from RtD studies. In HRI, however, we have not yet seen much work focusing on abstracting and elaborating intermediate-level designerly knowledge. Moreover, to the best of our knowledge, so far there is still no other HRI design work that adopts the usage of annotated portfolios on robotic artefacts \cite{Lupetti2021, APworkshop}. As pointed out by several scholars \cite{LuriaHRI21,workshop,Luria2019}, exploratory HRI design work carries crucial knowledge to understand the conceptual implications of robotic artifacts. To advance our understanding of designerly HRI work, investigations and examples are needed to foster and further the continuous questions and discussions around what design means to HRI.

To sum up, in HRI, where designerly approaches are not fully tapped yet, annotated portfolios can represent a particularly meaningful way to carry out the implicit concepts and assumptions embodied in robotic artefacts \cite{Lupetti2021}. To further the understanding of this approach, our work first demonstrates the usage of annotated portfolios for articulating designerly HRI knowledge by applying the approach on our two robotic artefacts with four different strategies. Subsequently, we look into how annotated portfolios, when performed by external experts, can help elicit new perspectives toward the same robotic artefacts.

\begin{figure*}[h]
 \begin{center}
  \includegraphics[width=1\textwidth]{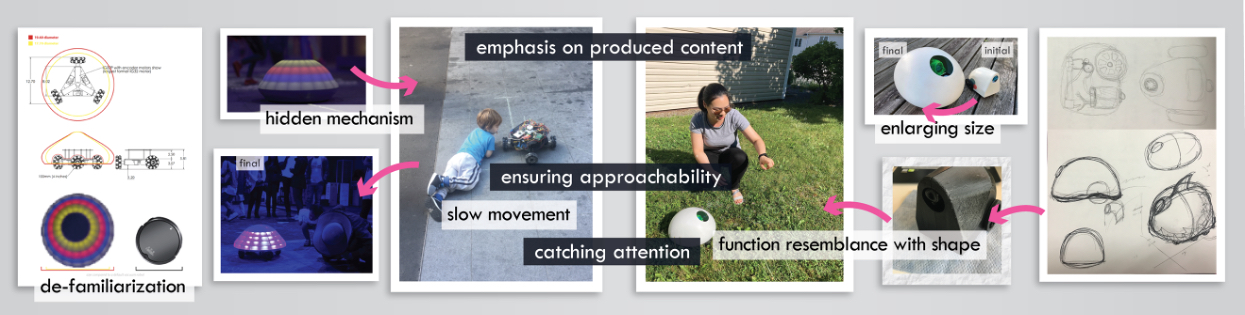}
  \end{center}
  \caption{Mapping design trajectories and iterations to capture conceptual similarities and re-examine design decisions.}
  \Description{Mapping}
  \label{fig:designer_ap2}
\end{figure*}

\section{Two Robotic Artefacts}
In the following, we briefly introduce two robotic artefacts from our RtD practice: BubbleBot and Woodie (see Figure \ref{fig:designer_ap1}). Both artefacts are comprised in our own annotated portfolio ('the designer's portfolio'), and later served as the exemplars for the annotated portfolios performed by external HRI experts ('the expert's portfolio'). While both design artefacts have a variety of similarities, including the deployment context, physical appearance, capabilities, and experiential qualities, they differ in regards to the designers' initial motivation and underlying research aims. Both artefacts have been designed and developed individually by each of the two first authors, both in the role of the lead designer, while the annotated portfolio has been created jointly by them. In order to understand how the portfolio presented in this paper elucidated new knowledge through the examination and comparison of the two cases, it is important to point out that neither of the designers or authors were aware of each other's projects at the time of implementing and publishing their respective research studies (see \cite{Lee2020, Hoggenmueller2020}).

\subsection{BubbleBot}

BubbleBot is a mobile robot carrying the function of bubble-blowing. Fast-paced contemporary life usually makes people miss out on wonderful moments \cite{Lee2020}. After observations in public spaces and embodied design workshops, the designers have applied the principles of ludic design \cite{gaver2002designing} and created BubbleBot: bursting bubbles at passers-by to invite for serendipitous interactions \cite{lee2019design}. With this project, the aim was to trigger conversations about the future roles and interaction paradigms of urban robots. The team deployed the initial design of BubbleBot in a populated common area of Cornell University in the US. The collected observation notes and video recordings were fed into the next design iteration of BubbleBot (see Figure \ref{fig:designer_ap2}), which will be deployed in the near future.

\subsection{Woodie}

Woodie is a slow-moving urban robot capable of drawing with conventional chalk sticks on the ground \cite{Hoggenmueller2020}. The overarching aim was to explore a novel form of pervasive urban display \cite{Hoggenmueller2019}, which produces content in a physicalised form. Building on previous research which highlights the experiential and transient qualities of non-digital displays \cite{Koeman2014}, the aim was to replicate and automate the same through a self-moving robotic platform. Woodie was built from scratch using electronic tinkering platforms (e.g. RaspberryPi, Arduino) and open-source software (e.g. Grbl). The design team deployed Woodie over three weeks in a quiet laneway in a densely populated northern suburb of Sydney, Australia, as part of an annual large-scale festival. During the deployment, data was collected through interviews, observation notes and video recordings.

\section{Designer's Portfolio}

In the following, we present our two cases represented in the form of an annotated portfolio. During the annotation process, we focused on various aspects of the robotic artefacts and captured different stages of the design processes. The annotated portfolio was created collaboratively by the two lead designers through the online whiteboard collaboration platform Miro\footnote{\url{https://miro.com/}, last accessed:
April 2021}.

\subsection{Annotating Interaction Qualities and Domain: The Ludic Urban Robot}

After getting familiar with each other's projects, we started to annotate two prominent images of each case, which depict the first in the wild deployment of the fully functional design artefacts (see Figure \ref{fig:designer_ap1}). Woodie was intended to use the ground as a large horizontal canvas by producing simple line drawings. Handing out chalk sticks to passers-by allowed them to directly manipulate the content, thus enabling tangible interactions. Woodie encouraged learning and creativity. For example, the design team observed children watching the robot’s drawings, and then copying or adapting them. BubbleBot was intended to elicit positive emotions and invite for serendipitous interactions by bursting bubbles at passers-by. Both robots have in common that they were oblivious of other people. Neither of them supported any form of direct input to take control over the robots’ behaviour (e.g. their movement). Instead their pure presence and actions fostered social interactions among people.

Considering the broader design domain, both projects illustrate the potential of augmenting public spaces through robotic artefacts. Compared to static and permanently deployed technologies (e.g. urban screens), they offer a lightweight solution to dynamically trigger playful and social interactions. While the robots differ in the degree of participation they enable, we argue that both act as placemaking facilitators in the sense they enhance urban experiences and promote a moment of happiness among passers-by, akin to buskers or street performers. Further, the absence of direct input controls resulted in people adopting a variety of approaches to engage with the robots. Coupled with their inherently playful activities, both artefacts manifest a proposition for ludic urban robotic interfaces.

\begin{figure*}[h]
  \includegraphics[width=1\textwidth]{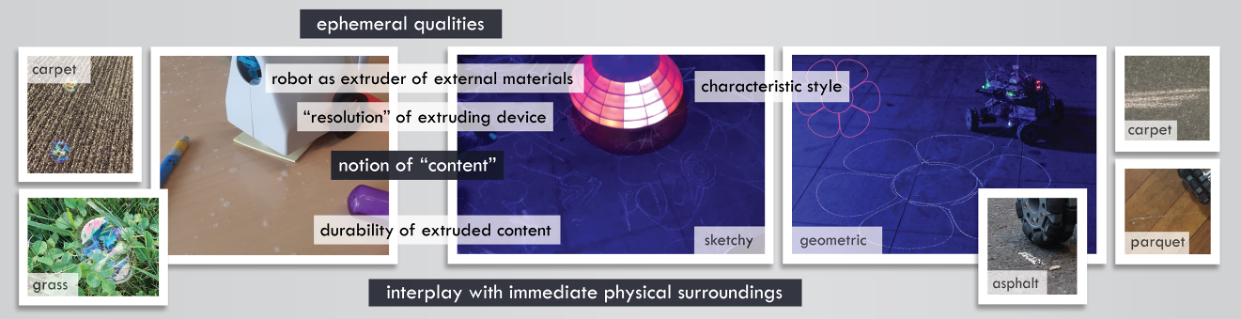}
  \caption{Zooming-in on the produced output to conceptualise the robots’ inherent characteristics and illustrate the interplay with the immediate surroundings.}
  \label{fig:designer_ap3}
\end{figure*}

\subsection{Mapping Design Trajectory and Iterations: The Robot’s Behaviour and Morphology}

In Figure \ref{fig:designer_ap2}, we captured some of the design considerations related to the morphology and behaviour of the robots, and how these evolved throughout the iterative design processes. For Woodie, the size was considered in reference to common domestic products (e.g. the vacuum cleaning robot Roomba). Following the ludic design principle of de-familiarization, a slightly larger size was set, which in turn also increased the chance of the robot being noticed by passers-by.

The initial design of BubbleBot was set with a smaller size to elicit a sense of friendliness, which, however, caused problems for people moving at a fast pace to notice the robot. Further, in both cases, the emphasis on the visibility of the produced “content” has influenced the final design decisions. For instance, the dimensions of Woodie had to be chosen so that they did not exceed the intended size of the drawings to ensure they were visible while Woodie was creating them. BubbleBot’s wheels were initially placed outside of its body, however later it was decided to hide all mechanical elements under the case to keep the main focus on the extruded bubbles. The same applies for Woodie, however, contrary as depicted on early renderings, for the final design, its shell was raised to allow people to observe the chalk stick. This decision was informed based on early tests, where the design team observed children sitting on the floor and being engaged by observing how the robot pulled the chalk stick behind. Further, those early tests revealed that children would come very close to the robot, which informed the decision for slow movements to ensure safety. For BubbleBot, the initial design was shaped as a cannon to reembody the robot’s activity of blowing bubbles. However, this was found to have an intimidating effect, which led to the new design with a round-shaped body resembling the form of soap bubbles.

\subsection{Zooming-In and Traversing: The Robot’s Output}

We decided in the next step to take a closer look at the “output” produced by the robots (see Figure \ref{fig:designer_ap3}). While both robots extrude external materials, interestingly, we found that the notion of “content” was a constant concern when designing Woodie but not so much for BubbleBot. This might be due to the difference of the resolution and the representational fidelity of the extruders: Woodie is able to draw simple iconic and symbolic representations, whereas BubbleBot either blows bubbles or not. In terms of the physical properties, the aspect of ephemerality (i.e. durability of the output) \cite{Doering2013} is more apparent with BubbleBot as bubbles disappear after a few seconds, whereas with Woodie the chalk drawings would stay for several hours or days depending on weather conditions, number of people walking through the space, and if content is overdrawn. Both the resolution of the extruders and the durability of the produced content also influence the experiences and engagement types observed during the deployments. BubbleBot invited people to stop by the robot and engage in short playful interactions with the bubbles less than a minute, while Woodie created longer-lasting engagements of people staying up to 20 minutes looking around the various drawings.

Based on these aspects, we looked into the earlier design phases and found that in both cases, the design teams tested the robots early on in various environments. In the case of BubbleBot, testings showed the consideration of the texture and humidity of grounds to withhold the surface tension of bubbles (e.g. carpet vs. grass). With Woodie, the testings were mostly concerned around the accuracy of the chalk drawings on various grounds, and which characteristic style would work best to avoid that the drawings look “imperfect” on rough terrain. Both findings indicate the deep interplay between the produced output with the immediate physical surroundings, and on a more abstract level that the consideration of the context plays an even more important role when designing cyber-physical artefacts, such as robots.

\begin{figure}[b]
  \includegraphics[width=\linewidth]{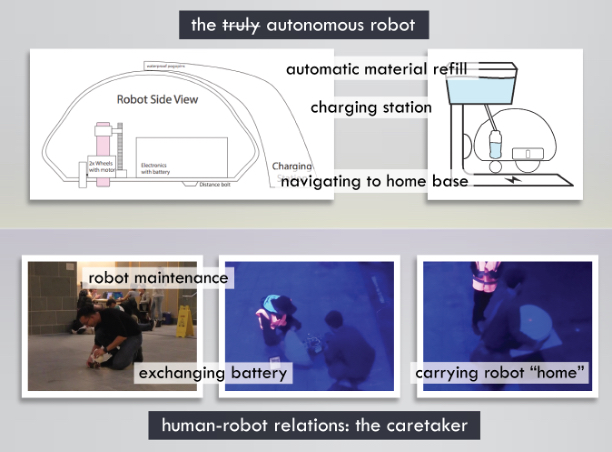}
  \caption{Early visions manifested through technical drawings juxtaposed with photographs showing the designers’ experiences during the actual deployment.}
  \Description{}
  \label{fig:designer_ap4}
\end{figure}

\subsection{Contrasting Ideals and Reality: The Truly Autonomous Robots}

We jointly reflected upon our early visions and ambitions regarding the level of autonomy that we intended to implement. For instance, in both cases we envisioned a base station for the robots, to which they would navigate autonomously for charging the batteries and refilling the extrusion materials. Figure \ref{fig:designer_ap4} captures some of those early visions inspired by existing products, such as Roomba, and robots in popular movies, such as Disney’s Baymax, who pops out from his re-charging container. We contrasted those visions with photographs taken during the actual deployments that capture our role and experiences as the robots’ caretakers. In both cases, we had to manually exchange batteries and renew the extruders. Due to the weight and size, the design team of Woodie created a purpose-built carrying apparatus, to which the robot could be fixated and brought back into a safe environment after each evening.

\section{Annotated Portfolio Workshops}
After having created our own annotated portfolio, we designed and deployed online workshops in which we invited experts from the HRI design field to create annotated portfolios on the same two robotic artefacts.

\subsection{Participants}
As a preliminary step to understand how annotated portfolios created by external performers can help elicit new perspectives, we aimed to get more insights by recruiting experts in the related field. The workshops were set with the eligibility statement that participants needed to be currently researching in HRI, and have followed or are interested in applying RtD approaches in their projects. We did not expect prior knowledge of annotated portfolios. Each participant was compensated with a US \$20 voucher. 
We recruited six experts through our university's mailing lists and social networks. All of our participants (4 females, 2 males) were currently enrolled in PhD-programs. Four resided in North America, two in Europe. Their research is situated in the broader area of social robotics and automation: two participants study social robotics for specific user groups, such as children (P4) and elders (P3, P4); two participants focus on accessibility (P1) and well-being (P5); the remaining participants are not addressing a specific target group or design domain (P1, P6). Three of the participants design their own robotic artefacts from scratch (P1) or modify the physical appearance of an existing robotic platform (P2, P6), while the remaining participants merely design the interactive behaviour of commercial robotic platforms. Three participants stated explicitly that they rely on RtD approaches in their projects (P1, P2, P6), two participants touch on RtD methods while pursuing a participatory design approach as their overarching research methodology (P4, P5). Three participants were familiar with the notion of annotated portfolios (P1, P2, P4), of which two read the relevant literature (P1, P2) and one plans to apply annotated portfolios in their own work (P2). The remaining participants (P3, P5, P6) were not familiar with the concept of annotated portfolios when commencing the study. 

\subsection{Procedure and Materials}
The six expert workshops were carried out individually (one expert per workshop) via the video conferencing platform Zoom with each workshop facilitated by both first authors and lasted sixty minutes in total. Before starting, we asked participants to open a prepared online whiteboard through the collaboration platform Miro and to share their screens. The participant information statement was outlined on the Miro board where we also asked for consent from the participants to audio and video record the workshops for later analysis. 

In the first step, we gave a 5-minute introduction about annotated portfolios. We showed participants an excerpt of Gaver's annotations of the Photostroller and Prayer Companion (see Figure 2 in \cite{Gaver2012}). We then introduced participants to our two own projects, Woodie and BubbleBot. We explained the main design considerations and rationale of each project, which were also represented in a textual form next to the two prominent images of each case (see Figure \ref{fig:designer_ap1}, \textit{without annotations}). Further, we showed participants a short 45-second video of each project, containing a compilation of video sequences of the in-the-wild deployments. In the next step, we asked participants to create two distinct annotated portfolios on the Miro board:\\

\noindent \textit{1) Annotating individual features:} here, participants were instructed to create unstructured annotations for each project individually. We did not put forward a specific annotation strategy yet as we wanted to understand how participants would intuitively make use of the approach, and which features (and insights) they find embodied in the artefacts and therefore worthwhile to communicate through an annotated portfolio.

\begin{figure*}[ht!]
  \includegraphics[width=\textwidth]{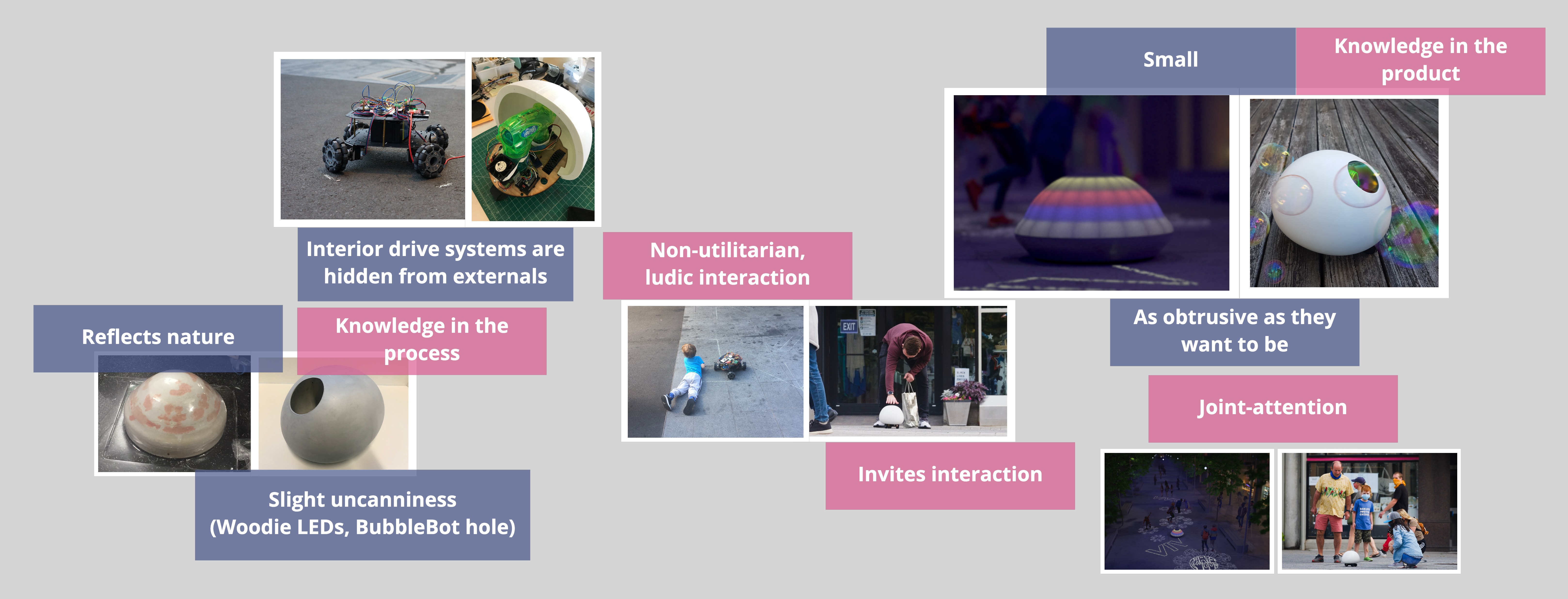}
  \caption{Excerpt of the annotated portfolio created by P1 in the expert workshop, showing annotations for Woodie and BubbleBot collectively.}
  \label{fig:expert_ap2}
\end{figure*}

\noindent \textit{2) Annotating artefacts collectively:} here, we instructed participants to focus on annotations concerning interaction qualities and design domain, and used Gaver's example again to explain this annotation strategy. Participants were asked to annotate both projects collectively, whereby annotations could be shared or applied to only one of the projects. An example can be found in Figure \ref{fig:expert_ap2}.\\

For both annotated portfolios, we prepared an image pool with 8 images per project from which participants could choose. The images were photographs taken by the researchers to depict the final artefacts, the in-the-wild deployments, as well as the unfinished prototypes during the making process. We left it open to participants to choose a single image or multiple images of each project. For the annotation process, we instructed participants to think out loud \cite{Lewis1993} and explain the annotations made. We aimed to receive more detailed information in addition to the rather visually organised portfolios and to contextualise the annotations with the participant's thinking process.
Further, participants were allowed to ask us questions about the artefacts, however, we made sure to answer them free from our own interpretations. After the annotation process, which lasted around 30-35 minutes in total, the study concluded with a 10-minute semi-structured interview with questions related to the overall experience of creating annotated portfolios. 

\subsection{Data Analysis}
We first reviewed the transcripts from each online workshop session separately and then conducted a thematic analysis \cite{BraCla06} to identify common themes across all data points (quotes). To get further insights on how the participants annotated the two projects, we reviewed the final portfolios and relevant video segments linked to the identified quotes. 
Two coders worked together to analyse the data, using a collaborative online whiteboard.

\section{Experts' Portfolios}
We here present the results from the annotated portfolios created by the robotic experts. The themes are presented in the chronological order of how their underlying annotations predominantly occurred in the course of the annotation process. There was a tendency that textual accounts made earlier in the annotation process were more on a surface level and of a descriptive nature relating to the robot's appearance. Later on participants also often added more conceptual annotations and such relating to the deployment context, the making process, and the broader methodological concerns.\footnote{Throughout this section, we quote textual accounts (i.e. annotations that participants wrote down on the Miro-board) using single quotation marks, and verbal accounts (e.g. statements made while thinking out loud and from the interviews) using double quotation marks and highlighted in \emph{italic}.}

\subsection{Appearance of the Robot with Underlying Assumptions} 
Annotations related to the appearance or shape of the robots were commonly brought up early in the annotation process. P1, for example, simply added that the robots seem `small' in relation to the situated urban context. Others emphasised the `abstract form' (P5), which P4 annotated as `whimsical'. Further, participants assigned morphological annotations, such as  `UFO dome' (P4) or `turtle' (P5) for Woodie and `smooth stone' (P1) or `cloud-like' (P2) for BubbleBot. Based on those features, some participants then made assumptions about the robot's behaviour and character traits, such as `pet-like' and `child-like' or being `naive and innocent' (P2, P4). P4 further elaborated on the shape of BubbleBot and added that it resembles bubbles which manifests the function of the robot. P6 added the `complexity of the robot is hidden behind a simple looking exterior', which could help users to make sense of the robots as being playful and nothing to be scared of. P5 added the annotation `no verbal interaction' and justified that she assumed this based on the provided video and photographs, but also that from a designer's perspective, she would tend towards adding implicit modes of interactions for a robot carrying a rather abstract shape.

\subsection{Deployment Context}
In addition to features related to the robot itself, the deployment context was reflected in the annotations of all participants, albeit emphasising different aspects. As both robots were depicted being deployed in public spaces, participants pointed out the `joint-attention' aroused by the robots, leading to 'emerging human-human engagement' between 'strangers'. Three participants indicated that no prior knowledge is required to engage with the robots as they are `universal' in the sense that \textit{``most people will see and understand [them]''}. Two participants further related to the `fast-paced context', and how the robots intend to reveal the `playful side' (P6) of people and help them to `be in the present' (P5). In this regard, P6 added another annotation which challenges this intentions in the sense that the designer expects `[...] certain safety and openness of the environment', but also that people within the environment  `[...] are open enough to interact with the robot'.

\subsection{Material Selection, Technical Implementation, \& Hacking Culture}
While all participants began by annotating either images showing close-ups of the robots or people interacting with the robots, two participants later on annotated images depicting the robot's internal components or illustrating different steps in the making process. P2 annotated an image of Woodie's unpolished shell after coming out of the vacuum former. She added that the `plastic indicates that the robots are envisioned for outdoor usage', implying that the deployment context influences the material selection. P1 made annotations related to the robots' drive system (i.e. `omni-directional' for Woodie, and `two-wheel drive' for BubbleBot). He elaborated that from a roboticist's perspective, this is important knowledge \textit{``to preserve for posterity''} as it provides insights on the selection of the drive system based on the robot's task or the behaviour to be implemented. Later, he added another image to his portfolio showing BubbleBot's interior with only a half-side shell covering the board on which the electrical components and the commercial bubble gun were attached to. He noted that this image illustrates well the process that robot designers, including himself, often follow in their practice: \textit{``[You use] what already existed, [...] and then you create this shell around it, physically and metaphorically. [...] So now this bubble blowing [gun] can move''}. In this regard, he also commented that designing robots means thinking in modules and layers, and further elaborated on the significance of the \textit{``hacking culture''} so that later \textit{``someone else will expand upon your robot design''}.

\subsection{Methodological Concerns} 
Later on in the annotation process, we observed how two participants moved forward from conceptual to adding more abstract annotations relating to key methodological concerns in design research. P1 further elaborated on the way how RtD approaches can produce knowledge in HRI. He added the annotation `knowledge in the process' next to the images depicting steps of the making process (shown in Fig. \ref{fig:expert_ap2}). He stated that `knowledge in the process' on the one hand provides insights about the technical implementation; further it \textit{``derives [...] something about the creators''} and \textit{``is a reflection of the creators' ideas''} by unfolding \textit{``the meaning about what they created [...] and how they were trying to do so''}. Later he added the annotation `knowledge in the product' and paraphrased what is commonly referred to as a design pattern \cite{Duyne2002} in the HCI literature: while pointing to an image depicting people gathering around the robot, he mentioned that \textit{``if you want to invite people to collaborate [with the robot]''}, then the robot \textit{``should be more flashy''} and \textit{``[you could] literally add lights to it''} (while pointing to a close-up of Woodie's low-resolution lighting display).

P3 referred to the three RtD approaches -- ``lab'', ``field'' and ``showroom'' -- described by Koskinen et al. \cite{Koskinen2011}. When juxtaposing the two artefacts, he added the annotation `showroom' for Woodie and `field' for BubbleBot. While acknowledging that he only vaguely remembers the concepts (indeed, he was the participant with the least touchpoints on RtD), he argued that the research objectives \textit{``feel''} more exploratory and speculative in the case of Woodie. Whereas, for BubbleBot, he could more clearly see that it was designed to investigate a \textit{``particular use of [social] robots''}. He interpreted BubbleBot as being more aligned with conventional ways of doing research in social robotics with a robot being designed for a for specific context (e.g. user's home), and its focus is \textit{``more oriented towards evaluation rather then exploration''}.

\section{Discussion}

As we have demonstrated throughout the paper, annotated portfolios can help generate conceptual themes from a collection of robotic artefacts and articulate designerly knowledge. We were able to gather various insights through the lens of both designer and experts. In our own annotations, we have highlighted the following features and design considerations for robotic artefacts:

\begin{itemize}[leftmargin=*]
    \item Focusing on \textit{interaction qualities} and \textit{domain aspects} \cite{Cullen2020}, we captured the stylistic similarities across the two robotic artefacts and portrayed a specific application area of urban robots acting as placemaking facilitators and enabling ludic interactions in cities. In addition, these annotations extrapolate the broader concern to free interactions with robots from the obnoxious habits of demanding constantly inputs and outputs from users, instead following the proposition of implicit urban interactions \cite{ju2015design}.
    \item By collectively \textit{mapping the design trajectory and iterations}, we pointed out the shared concerns and how they were addressed in each case individually over time. Three patterns of concern were discovered that influenced the design decisions in both cases: attracting people’s attention, ensuring approachability, and keeping emphasis on the robotic manipulation task and its outcome (i.e. visibility of the chalk drawings / bubbles).
    \item \textit{Zooming-in} on a specific aspect, in our case, the produced output, has helped us to bring out some of the inherent characteristics of each robotic device, and elaborate how those influenced passers-by engagement. Juxtaposing and conceptualising those characteristics could further lead to new design considerations: for example, in the case of BubbleBot, the encoding of implicit information through the size or intensity of the extruded bubbles, thus re-conceiving the robot as a producer of “content” similar to Woodie. Further, by traversing the various design stages, we captured the interplay between robot and environment, which required constant attention throughout the design process.
    \item By \textit{contrasting ideals and reality}, we unveiled some of the “dead ends” in the design process. Doing so helped us, as the designers, to understand our shared perceptions of a robot’s capabilities, which are shaped by existing products and also the worldview of the broader society towards robots. Contrasting with actual experiences, on the other hand, revealed some of the challenges when it comes to the permanent integration of robots into public space, thereby also considering alternative roles and worldviews on human-robot relations (e.g. humans as caretakers) \cite{Lupetti2019a}.
\end{itemize}

\noindent The expert annotations carried out additional conceptual themes that while being indexical to our cases, have the potential to be generalisable to other HRI designs and processes:

\begin{itemize}[leftmargin=*]
    \item Related to the deployment context in public space, the expert annotations pointed out the robots' function as a mediator for emerging `human-human interaction'.
    \item Several annotations emphasised how the robots' appearance and actions (e.g. `bubbles drawing attention') function as an opening encounter for interactions with passers-by. P1, further pointed out a \textit{design pattern} \cite{Duyne2002} that could be more widely applied to the design of urban robots concerning how to make people aware of a non-humanoid robot and initiate interactions through embedding low-resolution displays.
    \item We noticed annotations which resembled the form of \textit{criticism} \cite{Bardzell2010criticism}, which related to design research practices at large. For example, P1 mentioned \textit{``hacking culture''} as an influential factor for designerly approaches in the HRI community and also manifested in our robotic artefacts.
\end{itemize}

\subsection{Eliciting New Perspectives Through Experts' Annotations}

To understand how annotated portfolios performed by experts can bring in new perspectives on RtD studies, it has to be noted that the starting point for external observers is very different. This naturally affects the annotation process and the emerging knowledge generated. For example, at the beginning of the annotation process, participants, despite being experts in the field, often added textual accounts which resemble feedback that we obtained through previous user evaluations in the wild (e.g. related to the appearance of the robots and the tendency to anthropomorphise them). Indeed, participants sometimes even asked us if they shall add annotations from the perspective of a user, design researcher, or roboticist. It was therefore not our aim to compare the experts' annotations with the annotations from our own portfolio regarding the level of abstraction, nor to examine to what extent they exactly reflect our own initial designerly concerns and intents. As pointed out by L\"owgren, the \textit{``designer is in a privileged position to provide valuable abstractions''} due to first-hand information about the motivation of the work, the early design considerations, the making process, and the empirical evaluations \cite{Lowgren2013}. 
While acknowledging the missing familiarity with the annotated artefacts as a limitation, we want to emphasise how through analysing experts' annotations collectively and taking into account the underlying process itself, design researchers can obtain a form of creative and generative peer critique: 
\begin{itemize}[label={}, leftmargin=0pt]
    \item(1) In the simplest form, expert annotations can increase the saturation of an annotated portfolio. While L\"owgren points out that in the original proposition \cite{Gaver2012}, Gaver and Bowers increase the level of abstraction by extending the portfolio with additional artefacts \cite{Lowgren2013}, we argue that through multiple viewpoints, additional and previously not considered conceptual themes can be derived (e.g. the design pattern pointed out by P1 has not been previously considered in our own portfolio). 
    \item(2) Annotations made by expert's may carry the annotators' stance, which in turn can help the designer to reevaluate their design decisions in retrospect. For example, when P5 added the annotation `no verbal interaction', we made use of this as a prompt to elicit further insights on the reasoning, without confirming whether this was true for our cases or not. She argued then that her annotation was based on what she anticipated from the provided materials, but also personally \textit{``as a designer''}, she would argue that \textit{``the user should intuitively understand what type of interaction the robot is able to perform [...] based on the appearance''}.
    \item(3) Expert annotations can support designers with positioning their RtD cases to existing body of work in the field. Certainly, one can argue that the design researcher itself is an expert; however, in particular highly multi-disciplinary fields such as HRI can span a wide range of topics to which a RtD project might offer a valuable contribution even if not anticipated in the designer's initial framing. In our case, for example the proposition that both robots carry the function of a subtle opening encounter, which has been previously discussed for social robots in non-urban environments \cite{Hoffman2015}. 
    \item(4) Expert annotations can offer reflections on the contribution of RtD more broadly and trigger discussions of how to capture and disseminate designerly contributions. When P1 added the annotation `knowledge in the process' (see in Figure \ref{fig:expert_ap2}) and mentioned the role of \textit{``hacking cultures''}, he further elaborated that technical aspects in HRI studies are often not well documented, and therefore the potential to build on top of others' work is not fully tapped. This hints to previous endeavours in the larger RtD community: for example, the proposal by Desjardins et al. to disseminate designerly knowledge in the form of DIY tutorials \cite{Desjardins2017}, or considerations by Cul\'en et al. that \textit{``annotations can do more than pull towards concerns regarding abstraction''} \cite{Cullen2020}.
\end{itemize}

\subsection{Methodological Considerations for Annotated Portfolio Sessions with Experts }
Having discussed what to expect from annotated portfolios performed by experts and how this application brought in new perspectives on our own RtD cases, we here provide methodological considerations to help researchers and practitioners who consider to follow a similar approach for their own RtD practice. 
\begin{itemize}[label={}, leftmargin=0pt]
    \item (1) \textit{Selecting Participants:} In our case, we reached out to potential participants who conduct research within the field of HRI and were familiar with design research methods. This was decided as we assumed that participants who follow a similar research approach and conduct research in the same area could rather derive insights from our artefacts. Further, we assumed that design researchers can more easily apply such an approach, which was also confirmed by one participant who stated that \textit{``[this] is quite familiar territory''} to her (P5). On the other hand, the participant least familiar with RtD approaches (P3), provided interesting reflections on how the cases are positioned to existing HRI work that follows more conventional research methods. We therefore recommend that the choice of participants should be made based on the perspectives that researchers are seeking to bring in. Design researchers might more likely be able to articulate knowledge contributions from a RtD perspective, but also be more biased or in favour of the work by seeing through the lens of the same ``professional vision'' of their expertise \cite{Goodwin2015}. Participants from the broader research context on the other hand might bring in new perspectives and are able to position the work within the broader area.

    \item (2) \textit{Curating Image Pool:} The way we deployed annotated portfolios in the online workshops was highly image-heavy. We opted for a visual interpretation of annotated portfolios as we assumed that participants were more likely to be familiar with Gaver's visually organised proposition \cite{Gaver2012}. Further, we expected that it would be easier for participants to make sense of our projects within a restricted period of time. However, it has to be noted that the annotation process itself was the main foci of attention and that allowed experts to examine the robotic artefacts. Thus, the here chosen visual approach goes beyond other qualitative research methods, such as photo-elicitation interviews, where images are used as stimulus to evoke feelings, imagination, and thoughts \cite{tuomi2021spicing}. 
    Deploying a visually organised approach requires careful considerations for a) curating the image pool, and b) making sense of the annotations during the analysis. Providing multiple images and depicting various stages of the design process, in our case, led to a diverse range of outcomes as participants focused on different aspects in their annotation process. The content of the images also functioned as a filter and influenced the textual accounts. For example, P2 added the annotation `the weather is nice to interact with BubbleBot', thereby relating to the situational context depicted on a particular image. Often participants would also follow up with \textit{``at least from what I can see on the photo''} after adding a new annotation. Therefore, we argue that -- although as previously discussed one of the strengths of RtD is to explore robotic artefacts in context -- researchers have to carefully interpret the meaning of annotations when performed on contextualised images.

    \item (3) \textit{Deploying Annotation Strategies:} Given our image-heavy approach, in the beginning of the annotation process, participants often added annotations that would simply describe or comment on the images. While participants naturally began to start adding more abstract and conceptual annotations later in the process, deploying annotation strategies helped to force participants to juxtapose the artefacts and address broader domain concerns. However, on the other hand, participants also often stated that they found it more difficult to make those annotations. P3, for example, stated in regards to the example from Gaver that we provided in the beginning: \textit{``I feel like these phrases are rather abstract [...] I couldn't relate to it.''} P2 and P5 also referred to the example as `researcher/designer language' and mentioned that it would be hard to understand without further debriefing.

    \item (4) \textit{Think-aloud Protocol:} Throughout the annotation process, we asked participants to think out loud \cite{Lewis1993} to elaborate on their annotations and for us to further understand how they made sense of the projects. The information we gathered enabled further data analysis on linking the annotations made and logic behind their inductive reasoning process. Further, we observed that expressing their thoughts often helped participants to refine annotations and even led them toward new ones. 

    \item (5) \textit{Facilitating Communication:} Additionally to applying the think aloud protocol, facilitating communication between researchers and participants was important. At the beginning participants were often seeking for confirmation if their annotations made sense or if they are \textit{``completely wrong''}. Therefore, it was important to emphasise repeatedly the openness of the approach and that the artefacts can be annotated in various ways, also in terms of the level of abstraction. Further, participants often asked questions about the projects given the limited account of background knowledge that the provided materials would give. In this regard, however, the challenge is to answer the questions \textit{not} in an interpretive manner to avoid biasing participants in their annotation process.
\end{itemize}

\subsection{Annotated Portfolios in the Designerly HRI Toolbox}

Despite the growing interest on design research in the HRI community \cite{LuriaHRI21}, most of the RtD artefacts are being published as individual case studies \cite{Overgoor2018IdleBot,Gamboa2021Drones,Hoggenmueller2020,Lee2020,Hoffman2015,AndersonBashan2018}, and methods and tools mostly investigate tailored prototyping approaches \cite{Zamfirescu2021} and design processes \cite{Oliveira2021}. Lupetti et al. recently called for more conceptual investigations, arguing that singular robotic artefacts are addressing very specific and unique research problems \cite{Lupetti2021}. They proposed to contextualise intermediate-level knowledge in HRI with annotated portfolios, amongst others, to reflect on design decisions, understand conceptual implications, and document and present generated HRI knowledge collectively.

In our work, we exemplified how annotated portfolios can be applied in the context of HRI. Through our own annotated portfolio, we in particular investigated various new annotation strategies to bring upfront the conceptual implications of our research artefacts both individually and collectively. While annotated portfolios in HCI have been predominantly applied to finished artefacts, we addressed aspects and perspectives beyond the artefact by including visual documents from the actual making process. This is in line with the initial proposition of annotated portfolios being open to interpretation and appropriation \cite{Gaver2012}, and addresses concerns in HRI that robots as physical artefacts hold knowledge in the iterative process of making, which is often poorly documented \cite{LuriaHRI21}. For example, through the strategy of \textit{mapping the design trajectory and iterations}, we revisited past design decisions. The documentation and presentation can offer other researchers insights on what factors to consider for the design of social urban robots and how those factors might influence the user experience. Going even further back in the process by considering early visions and contrast with the final implementation can not only further provide HRI researchers and practitioners with genuine insights to inform future designs, but also to encourage and support rethinking preconceptions and roles attributed to robots. \textit{Zoom\-ing-in} allows the designer to carve-out a specific aspect of the interface and interaction, which is important as physical robots are often not limited to a single user interface and are entangled in complex and changing contexts \cite{Lupetti2021}.

With respect to expert annotations as an additional variation, we argue that their analysis can provide HRI designers with a new frame of references: how their particular design instances are adding to and how to position them to existing HRI knowledge \cite{Lupetti2021}. Moreover, our approach also indicates that creating annotated portfolios with experts can serve as a method to discuss and critique the contributions and dissemination of RtD projects in HRI. In their analysis and critique on RtD approaches in HCI, Zimmerman et al. reported that several \textit{``canonical examples of design research''} have been repeatedly cited in expert interviews to elaborate on the contributions of design research to the larger community \cite{Zimmerman2010}. In HRI where those historical and canonical examples do not widely exist yet, our approach could enable an open discussion on the contributions and limitations of designerly HRI approaches through the in-depth examination of provisional and contingent RtD projects.

While we argue that annotated portfolios can be a promising approach within the HRI design toolbox, there are certain limitations which need further considerations. For instance, to what extent can phenomenological aspects (i.e. user’s perceptions and emotional responses), as well as embodied interactions and user trajectories over time be captured? In a similar vein, one of the experts, P1, brought forward the idea of video-based portfolios to also capture the \textit{``timing of the movements''}. Further, for the design of embodied robots there are often additional modalities at play, such as speech, sound and tactile feedback, that can not be easily captured visually. In this case, additional visual abstractions are needed (i.e. diagrams, sketches or storyboards) to capture the work in the form of a portfolio, and to reveal some of the tacit knowledge.

\section{Conclusion}
Annotated portfolios -- as part of the intermediate-level knowledge toolbox -- represent a unique approach to articulate designerly knowledge in a more abstract form while retaining indexical connection with the produced artefacts. While annotated portfolios have been widely used in the HCI design community, there are related subfields, such as HRI, where the approach has not been fully addressed yet. In this paper, we analysed two of our own RtD studies on urban robots by building on the approach of annotated portfolios and demonstrate how annotated portfolios can help elicit new perspectives when performed by experts.

We presented our own portfolio as well as results from the expert workshops, and from that, we offered reflections valuable for researchers who are conducting designerly HRI work. Moreover, we elaborated on how annotated portfolios -- when performed by other experts -- can elicit new perspectives on RtD studies in the form of a creative and generative peer critique. Finally, we made a case for adding annotated portfolios to the designerly HRI toolbox. We suggest that this approach can serve as a legitimate form of inquiry to congregate design work and advance the understanding of what design means to HRI. Beyond HRI, the proposed approach, can be applied to other technology- and science-driven domains, where RtD studies might add valuable perspectives, such as HCI for space exploration \cite{spaceworkshop} or biodesigned systems \cite{Gough2020}. 

\begin{acks}
We thank all the participants for taking part in this research. Further, we would like to thank the organisers and participants of the ``First international workshop on Designerly HRI Knowledge'' at RO-MAN 2020. We acknowledge that their feedback and discussions have further inspired this work, which is partially based on original ideas that we published in a workshop position paper and presented at the workshop. We also thank the anonymous DIS’21 reviewers and ACs for their constructive feedback and suggestions how to make this contribution stronger.
\end{acks}

\balance{}
\balance{}

\bibliographystyle{ACM-Reference-Format}
\bibliography{sample-base}


\begin{thebibliography}{55}


\ifx \showCODEN    \undefined \def \showCODEN     #1{\unskip}     \fi
\ifx \showDOI      \undefined \def \showDOI       #1{#1}\fi
\ifx \showISBNx    \undefined \def \showISBNx     #1{\unskip}     \fi
\ifx \showISBNxiii \undefined \def \showISBNxiii  #1{\unskip}     \fi
\ifx \showISSN     \undefined \def \showISSN      #1{\unskip}     \fi
\ifx \showLCCN     \undefined \def \showLCCN      #1{\unskip}     \fi
\ifx \shownote     \undefined \def \shownote      #1{#1}          \fi
\ifx \showarticletitle \undefined \def \showarticletitle #1{#1}   \fi
\ifx \showURL      \undefined \def \showURL       {\relax}        \fi
\providecommand\bibfield[2]{#2}
\providecommand\bibinfo[2]{#2}
\providecommand\natexlab[1]{#1}
\providecommand\showeprint[2][]{arXiv:#2}

\bibitem[Alves-Oliveira et~al\mbox{.}(2021)]%
        {Oliveira2021}
\bibfield{author}{\bibinfo{person}{Patr\'{\i}cia Alves-Oliveira}, \bibinfo{person}{Patr\'{\i}cia Arriaga}, \bibinfo{person}{Ana Paiva}, {and} \bibinfo{person}{Guy Hoffman}.} \bibinfo{year}{2021}\natexlab{}.
\newblock \showarticletitle{Children as Robot Designers}. In \bibinfo{booktitle}{\emph{Proceedings of the 2021 ACM/IEEE International Conference on Human-Robot Interaction}} (Boulder, CO, USA) \emph{(\bibinfo{series}{HRI '21})}. \bibinfo{publisher}{Association for Computing Machinery}, \bibinfo{address}{New York, NY, USA}, \bibinfo{pages}{399–408}.
\newblock
\showISBNx{9781450382892}
\urldef\tempurl%
\url{https://doi.org/10.1145/3434073.3444650}
\showDOI{\tempurl}


\bibitem[Anderson-Bashan et~al\mbox{.}(2018)]%
        {AndersonBashan2018}
\bibfield{author}{\bibinfo{person}{Lucy Anderson-Bashan}, \bibinfo{person}{Benny Megidish}, \bibinfo{person}{Hadas Erel}, \bibinfo{person}{Iddo Wald}, \bibinfo{person}{Guy Hoffman}, \bibinfo{person}{Oren Zuckerman}, {and} \bibinfo{person}{Andrey Grishko}.} \bibinfo{year}{2018}\natexlab{}.
\newblock \showarticletitle{The Greeting Machine: An Abstract Robotic Object for Opening Encounters}. In \bibinfo{booktitle}{\emph{2018 27th IEEE International Symposium on Robot and Human Interactive Communication (RO-MAN)}}. \bibinfo{pages}{595--602}.
\newblock
\urldef\tempurl%
\url{https://doi.org/10.1109/ROMAN.2018.8525516}
\showDOI{\tempurl}


\bibitem[Bardzell et~al\mbox{.}(2016)]%
        {Bardzell2016}
\bibfield{author}{\bibinfo{person}{Jeffrey Bardzell}, \bibinfo{person}{Shaowen Bardzell}, \bibinfo{person}{Peter Dalsgaard}, \bibinfo{person}{Shad Gross}, {and} \bibinfo{person}{Kim Halskov}.} \bibinfo{year}{2016}\natexlab{}.
\newblock \showarticletitle{Documenting the Research Through Design Process}. In \bibinfo{booktitle}{\emph{Proceedings of the 2016 ACM Conference on Designing Interactive Systems}} (Brisbane, QLD, Australia) \emph{(\bibinfo{series}{DIS '16})}. \bibinfo{publisher}{Association for Computing Machinery}, \bibinfo{address}{New York, NY, USA}, \bibinfo{pages}{96–107}.
\newblock
\showISBNx{9781450340311}
\urldef\tempurl%
\url{https://doi.org/10.1145/2901790.2901859}
\showDOI{\tempurl}


\bibitem[Bardzell et~al\mbox{.}(2010)]%
        {Bardzell2010criticism}
\bibfield{author}{\bibinfo{person}{Jeffrey Bardzell}, \bibinfo{person}{Jay Bolter}, {and} \bibinfo{person}{Jonas L\"{o}wgren}.} \bibinfo{year}{2010}\natexlab{}.
\newblock \showarticletitle{Interaction Criticism: Three Readings of an Interaction Design, and What They Get Us}.
\newblock \bibinfo{journal}{\emph{Interactions}} \bibinfo{volume}{17}, \bibinfo{number}{2} (\bibinfo{date}{March} \bibinfo{year}{2010}), \bibinfo{pages}{32–37}.
\newblock
\showISSN{1072-5520}
\urldef\tempurl%
\url{https://doi.org/10.1145/1699775.1699783}
\showDOI{\tempurl}


\bibitem[Bowers(2012)]%
        {bowers2012logic}
\bibfield{author}{\bibinfo{person}{John Bowers}.} \bibinfo{year}{2012}\natexlab{}.
\newblock \showarticletitle{The Logic of Annotated Portfolios: Communicating the Value of 'Research through Design'}. In \bibinfo{booktitle}{\emph{Proceedings of the Designing Interactive Systems Conference}} (Newcastle Upon Tyne, United Kingdom) \emph{(\bibinfo{series}{DIS '12})}. \bibinfo{publisher}{Association for Computing Machinery}, \bibinfo{address}{New York, NY, USA}, \bibinfo{pages}{68–77}.
\newblock
\showISBNx{9781450312103}
\urldef\tempurl%
\url{https://doi.org/10.1145/2317956.2317968}
\showDOI{\tempurl}


\bibitem[Braun and Clarke(2006)]%
        {BraCla06}
\bibfield{author}{\bibinfo{person}{Virginia Braun} {and} \bibinfo{person}{Victoria Clarke}.} \bibinfo{year}{2006}\natexlab{}.
\newblock \showarticletitle{Using thematic analysis in psychology}.
\newblock \bibinfo{journal}{\emph{Qualitative Research in Psychology}} \bibinfo{volume}{3}, \bibinfo{number}{2} (\bibinfo{year}{2006}), \bibinfo{pages}{77--101}.
\newblock


\bibitem[Cheon and Su(2018)]%
        {Cheon2018}
\bibfield{author}{\bibinfo{person}{EunJeong Cheon} {and} \bibinfo{person}{Norman~Makoto Su}.} \bibinfo{year}{2018}\natexlab{}.
\newblock \showarticletitle{Futuristic Autobiographies: Weaving Participant Narratives to Elicit Values around Robots}. In \bibinfo{booktitle}{\emph{Proceedings of the 2018 ACM/IEEE International Conference on Human-Robot Interaction}} (Chicago, IL, USA) \emph{(\bibinfo{series}{HRI '18})}. \bibinfo{publisher}{Association for Computing Machinery}, \bibinfo{address}{New York, NY, USA}, \bibinfo{pages}{388–397}.
\newblock
\showISBNx{9781450349536}
\urldef\tempurl%
\url{https://doi.org/10.1145/3171221.3171244}
\showDOI{\tempurl}


\bibitem[Cross(1982)]%
        {cross1982designerly}
\bibfield{author}{\bibinfo{person}{Nigel Cross}.} \bibinfo{year}{1982}\natexlab{}.
\newblock \showarticletitle{Designerly ways of knowing}.
\newblock \bibinfo{journal}{\emph{Design Studies}} \bibinfo{volume}{3}, \bibinfo{number}{4} (\bibinfo{year}{1982}), \bibinfo{pages}{221 -- 227}.
\newblock
\showISSN{0142-694X}
\urldef\tempurl%
\url{https://doi.org/10.1016/0142-694X(82)90040-0}
\showDOI{\tempurl}
\newblock
\shownote{Special Issue Design Education}.


\bibitem[Cul\'{e}n et~al\mbox{.}(2020)]%
        {Cullen2020}
\bibfield{author}{\bibinfo{person}{Alma~Leora Cul\'{e}n}, \bibinfo{person}{Jorun B\o{}rsting}, {and} \bibinfo{person}{William Gaver}.} \bibinfo{year}{2020}\natexlab{}.
\newblock \showarticletitle{Strategies for Annotating Portfolios: Mapping Designs for New Domains}. In \bibinfo{booktitle}{\emph{Proceedings of the 2020 ACM Designing Interactive Systems Conference}} (Eindhoven, Netherlands) \emph{(\bibinfo{series}{DIS ’20})}. \bibinfo{publisher}{Association for Computing Machinery}, \bibinfo{address}{New York, NY, USA}, \bibinfo{pages}{1633–1645}.
\newblock
\showISBNx{9781450369749}
\urldef\tempurl%
\url{https://doi.org/10.1145/3357236.3395490}
\showDOI{\tempurl}


\bibitem[Dalsgaard and Dindler(2014)]%
        {Dalsgaard2014}
\bibfield{author}{\bibinfo{person}{Peter Dalsgaard} {and} \bibinfo{person}{Christian Dindler}.} \bibinfo{year}{2014}\natexlab{}.
\newblock \showarticletitle{Between Theory and Practice: Bridging Concepts in HCI Research}. In \bibinfo{booktitle}{\emph{Proceedings of the SIGCHI Conference on Human Factors in Computing Systems}} (Toronto, Ontario, Canada) \emph{(\bibinfo{series}{CHI '14})}. \bibinfo{publisher}{Association for Computing Machinery}, \bibinfo{address}{New York, NY, USA}, \bibinfo{pages}{1635–1644}.
\newblock
\showISBNx{9781450324731}
\urldef\tempurl%
\url{https://doi.org/10.1145/2556288.2557342}
\showDOI{\tempurl}


\bibitem[Desjardins et~al\mbox{.}(2017)]%
        {Desjardins2017}
\bibfield{author}{\bibinfo{person}{Audrey Desjardins}, \bibinfo{person}{Ron Wakkary}, \bibinfo{person}{Will Odom}, \bibinfo{person}{Henry Lin}, {and} \bibinfo{person}{Markus~Lorenz Schilling}.} \bibinfo{year}{2017}\natexlab{}.
\newblock \showarticletitle{Exploring DIY Tutorials as a Way to Disseminate Research through Design}.
\newblock \bibinfo{journal}{\emph{Interactions}} \bibinfo{volume}{24}, \bibinfo{number}{4} (\bibinfo{date}{June} \bibinfo{year}{2017}), \bibinfo{pages}{78–82}.
\newblock
\showISSN{1072-5520}
\urldef\tempurl%
\url{https://doi.org/10.1145/3098319}
\showDOI{\tempurl}


\bibitem[D\"{o}ring et~al\mbox{.}(2013)]%
        {Doering2013}
\bibfield{author}{\bibinfo{person}{Tanja D\"{o}ring}, \bibinfo{person}{Axel Sylvester}, {and} \bibinfo{person}{Albrecht Schmidt}.} \bibinfo{year}{2013}\natexlab{}.
\newblock \showarticletitle{Ephemeral User Interfaces: Valuing the Aesthetics of Interface Components That Do Not Last}.
\newblock \bibinfo{journal}{\emph{Interactions}} \bibinfo{volume}{20}, \bibinfo{number}{4} (\bibinfo{date}{7} \bibinfo{year}{2013}), \bibinfo{pages}{32–37}.
\newblock
\showISSN{1072-5520}
\urldef\tempurl%
\url{https://doi.org/10.1145/2486227.2486235}
\showDOI{\tempurl}


\bibitem[Dunne and Raby(2007)]%
        {Dunne2007}
\bibfield{author}{\bibinfo{person}{Anthony Dunne} {and} \bibinfo{person}{Fiona Raby}.} \bibinfo{year}{2007}\natexlab{}.
\newblock \bibinfo{title}{Technological Dreams Series: No.1, Robots}.
\newblock \bibinfo{howpublished}{\url{http://dunneandraby.co.uk/content/projects/10/0}, \textit{last accessed: June 2020}}.
\newblock


\bibitem[Duyne et~al\mbox{.}(2002)]%
        {Duyne2002}
\bibfield{author}{\bibinfo{person}{Douglas K.~Van Duyne}, \bibinfo{person}{James Landay}, {and} \bibinfo{person}{Jason~I. Hong}.} \bibinfo{year}{2002}\natexlab{}.
\newblock \bibinfo{booktitle}{\emph{The Design of Sites: Patterns, Principles, and Processes for Crafting a Customer-Centered Web Experience}}.
\newblock \bibinfo{publisher}{Addison-Wesley Longman Publishing Co., Inc.}, \bibinfo{address}{USA}.
\newblock
\showISBNx{020172149X}


\bibitem[Fallman and Stolterman(2010)]%
        {Fallman2010}
\bibfield{author}{\bibinfo{person}{Daniel Fallman} {and} \bibinfo{person}{Erik Stolterman}.} \bibinfo{year}{2010}\natexlab{}.
\newblock \showarticletitle{Establishing criteria of rigour and relevance in interaction design research}.
\newblock \bibinfo{journal}{\emph{Digital Creativity}} \bibinfo{volume}{21}, \bibinfo{number}{4} (\bibinfo{year}{2010}), \bibinfo{pages}{265--272}.
\newblock
\urldef\tempurl%
\url{https://doi.org/10.1080/14626268.2010.548869}
\showDOI{\tempurl}
\showeprint{https://doi.org/10.1080/14626268.2010.548869}


\bibitem[Gamboa et~al\mbox{.}(2021)]%
        {Gamboa2021Drones}
\bibfield{author}{\bibinfo{person}{Mafalda Gamboa}, \bibinfo{person}{Mohammad Obaid}, {and} \bibinfo{person}{Sara Ljungblad}.} \bibinfo{year}{2021}\natexlab{}.
\newblock \showarticletitle{Ritual Drones: Designing and Studying Critical Flying Companions}. In \bibinfo{booktitle}{\emph{Companion of the 2021 ACM/IEEE International Conference on Human-Robot Interaction}} (Boulder, CO, USA) \emph{(\bibinfo{series}{HRI '21 Companion})}. \bibinfo{publisher}{Association for Computing Machinery}, \bibinfo{address}{New York, NY, USA}, \bibinfo{pages}{562–564}.
\newblock
\showISBNx{9781450382908}
\urldef\tempurl%
\url{https://doi.org/10.1145/3434074.3446363}
\showDOI{\tempurl}


\bibitem[Gaver and Bowers(2012)]%
        {Gaver2012}
\bibfield{author}{\bibinfo{person}{Bill Gaver} {and} \bibinfo{person}{John Bowers}.} \bibinfo{year}{2012}\natexlab{}.
\newblock \showarticletitle{Annotated Portfolios}.
\newblock \bibinfo{journal}{\emph{Interactions}} \bibinfo{volume}{19}, \bibinfo{number}{4} (\bibinfo{date}{7} \bibinfo{year}{2012}), \bibinfo{pages}{40–49}.
\newblock
\showISSN{1072-5520}
\urldef\tempurl%
\url{https://doi.org/10.1145/2212877.2212889}
\showDOI{\tempurl}


\bibitem[Gaver(2002)]%
        {gaver2002designing}
\bibfield{author}{\bibinfo{person}{William Gaver}.} \bibinfo{year}{2002}\natexlab{}.
\newblock \showarticletitle{Designing for homo ludens}.
\newblock \bibinfo{journal}{\emph{I3 Magazine}}  \bibinfo{volume}{12} (\bibinfo{year}{2002}), \bibinfo{pages}{2--6}.
\newblock


\bibitem[Gaver(2012)]%
        {gaver2012should}
\bibfield{author}{\bibinfo{person}{William Gaver}.} \bibinfo{year}{2012}\natexlab{}.
\newblock \showarticletitle{What should we expect from research through design?}. In \bibinfo{booktitle}{\emph{Proceedings of the SIGCHI conference on human factors in computing systems}}. \bibinfo{pages}{937--946}.
\newblock


\bibitem[Goodwin(2015)]%
        {Goodwin2015}
\bibfield{author}{\bibinfo{person}{Charles Goodwin}.} \bibinfo{year}{2015}\natexlab{}.
\newblock \bibinfo{booktitle}{\emph{Professional Vision}}.
\newblock \bibinfo{publisher}{Springer Fachmedien Wiesbaden}, \bibinfo{address}{Wiesbaden}, \bibinfo{pages}{387--425}.
\newblock
\showISBNx{978-3-531-19381-6}
\urldef\tempurl%
\url{https://doi.org/10.1007/978-3-531-19381-6_20}
\showDOI{\tempurl}


\bibitem[Gough et~al\mbox{.}(2020)]%
        {Gough2020}
\bibfield{author}{\bibinfo{person}{Phillip Gough}, \bibinfo{person}{Larissa Pschetz}, \bibinfo{person}{Naseem Ahmadpour}, \bibinfo{person}{Leigh-Anne Hepburn}, \bibinfo{person}{Clare Cooper}, \bibinfo{person}{Carolina Ramirez-Figueroa}, {and} \bibinfo{person}{Oron Catts}.} \bibinfo{year}{2020}\natexlab{}.
\newblock \showarticletitle{The Nature of Biodesigned Systems: Directions for HCI}. In \bibinfo{booktitle}{\emph{Companion Publication of the 2020 ACM Designing Interactive Systems Conference}} (Eindhoven, Netherlands) \emph{(\bibinfo{series}{DIS' 20 Companion})}. \bibinfo{publisher}{Association for Computing Machinery}, \bibinfo{address}{New York, NY, USA}, \bibinfo{pages}{389–392}.
\newblock
\showISBNx{9781450379878}
\urldef\tempurl%
\url{https://doi.org/10.1145/3393914.3395908}
\showDOI{\tempurl}


\bibitem[Hauser et~al\mbox{.}(2018)]%
        {hauser2018annotated}
\bibfield{author}{\bibinfo{person}{Sabrina Hauser}, \bibinfo{person}{Doenja Oogjes}, \bibinfo{person}{Ron Wakkary}, {and} \bibinfo{person}{Peter-Paul Verbeek}.} \bibinfo{year}{2018}\natexlab{}.
\newblock \showarticletitle{An Annotated Portfolio on Doing Postphenomenology Through Research Products}. In \bibinfo{booktitle}{\emph{Proceedings of the 2018 Designing Interactive Systems Conference}} (Hong Kong, China) \emph{(\bibinfo{series}{DIS '18})}. \bibinfo{publisher}{Association for Computing Machinery}, \bibinfo{address}{New York, NY, USA}, \bibinfo{pages}{459–471}.
\newblock
\showISBNx{9781450351980}
\urldef\tempurl%
\url{https://doi.org/10.1145/3196709.3196745}
\showDOI{\tempurl}


\bibitem[Hoang et~al\mbox{.}(2018)]%
        {Hoang2018what}
\bibfield{author}{\bibinfo{person}{Ti Hoang}, \bibinfo{person}{Rohit~Ashok Khot}, \bibinfo{person}{Noel Waite}, {and} \bibinfo{person}{Florian~'Floyd' Mueller}.} \bibinfo{year}{2018}\natexlab{}.
\newblock \showarticletitle{What Can Speculative Design Teach Us about Designing for Healthcare Services?}. In \bibinfo{booktitle}{\emph{Proceedings of the 30th Australian Conference on Computer-Human Interaction}} (Melbourne, Australia) \emph{(\bibinfo{series}{OzCHI '18})}. \bibinfo{publisher}{Association for Computing Machinery}, \bibinfo{address}{New York, NY, USA}, \bibinfo{pages}{463–472}.
\newblock
\showISBNx{9781450361880}
\urldef\tempurl%
\url{https://doi.org/10.1145/3292147.3292160}
\showDOI{\tempurl}


\bibitem[Hoffman et~al\mbox{.}(2015)]%
        {Hoffman2015}
\bibfield{author}{\bibinfo{person}{Guy Hoffman}, \bibinfo{person}{Oren Zuckerman}, \bibinfo{person}{Gilad Hirschberger}, \bibinfo{person}{Michal Luria}, {and} \bibinfo{person}{Tal Shani-Sherman}.} \bibinfo{year}{2015}\natexlab{}.
\newblock \showarticletitle{Design and Evaluation of a Peripheral Robotic Conversation Companion}. In \bibinfo{booktitle}{\emph{2015 10th ACM/IEEE International Conference on Human-Robot Interaction (HRI)}}. \bibinfo{pages}{3--10}.
\newblock


\bibitem[Hoggenmueller et~al\mbox{.}(2020a)]%
        {Hoggenmueller2020}
\bibfield{author}{\bibinfo{person}{Marius Hoggenmueller}, \bibinfo{person}{Luke Hespanhol}, {and} \bibinfo{person}{Martin Tomitsch}.} \bibinfo{year}{2020}\natexlab{a}.
\newblock \showarticletitle{Stop and Smell the Chalk Flowers: A Robotic Probe for Investigating Urban Interaction with Physicalised Displays}. In \bibinfo{booktitle}{\emph{Proceedings of the 2020 CHI Conference on Human Factors in Computing Systems}} (Honolulu, HI, USA) \emph{(\bibinfo{series}{CHI ’20})}. \bibinfo{publisher}{Association for Computing Machinery}, \bibinfo{address}{New York, NY, USA}, \bibinfo{pages}{1–14}.
\newblock
\showISBNx{9781450367080}
\urldef\tempurl%
\url{https://doi.org/10.1145/3313831.3376676}
\showDOI{\tempurl}


\bibitem[Hoggenmueller et~al\mbox{.}(2019)]%
        {Hoggenmueller2019}
\bibfield{author}{\bibinfo{person}{Marius Hoggenmueller}, \bibinfo{person}{Luke Hespanhol}, \bibinfo{person}{Alexander Wiethoff}, {and} \bibinfo{person}{Martin Tomitsch}.} \bibinfo{year}{2019}\natexlab{}.
\newblock \showarticletitle{Self-moving Robots and Pulverized Urban Displays: Newcomers in the Pervasive Display Taxonomy}. In \bibinfo{booktitle}{\emph{Proceedings of the 8th ACM International Symposium on Pervasive Displays}} (Palermo, Italy) \emph{(\bibinfo{series}{PerDis '19})}. \bibinfo{publisher}{ACM}, \bibinfo{address}{New York, NY, USA}, Article \bibinfo{articleno}{1}, \bibinfo{numpages}{8}~pages.
\newblock
\showISBNx{978-1-4503-6751-6}
\urldef\tempurl%
\url{https://doi.org/10.1145/3321335.3324950}
\showDOI{\tempurl}


\bibitem[Hoggenmueller et~al\mbox{.}(2020b)]%
        {APworkshop}
\bibfield{author}{\bibinfo{person}{Marius Hoggenmueller}, \bibinfo{person}{Wen-Ying Lee}, \bibinfo{person}{Luke Hespanhol}, \bibinfo{person}{Martin Tomitsch}, {and} \bibinfo{person}{Malte Jung}.} \bibinfo{year}{2020}\natexlab{b}.
\newblock \bibinfo{title}{Beyond the Robotic Artefact: Capturing Designerly HRI Knowledge through Annotated Portfolios}.  (\bibinfo{year}{2020}).
\newblock
\newblock
\shownote{Position paper presented at the 1st First international workshop on Designerly HRI Knowledge. Held in conjunction with the 29th IEEE International Conference on Robot and Human Interactive Communication (RO-MAN’20)}.


\bibitem[H\"{o}\"{o}k et~al\mbox{.}(2015a)]%
        {Hook2015framing}
\bibfield{author}{\bibinfo{person}{Kristina H\"{o}\"{o}k}, \bibinfo{person}{Jeffrey Bardzell}, \bibinfo{person}{Simon Bowen}, \bibinfo{person}{Peter Dalsgaard}, \bibinfo{person}{Stuart Reeves}, {and} \bibinfo{person}{Annika Waern}.} \bibinfo{year}{2015}\natexlab{a}.
\newblock \showarticletitle{Framing IxD Knowledge}.
\newblock \bibinfo{journal}{\emph{Interactions}} \bibinfo{volume}{22}, \bibinfo{number}{6} (\bibinfo{date}{Oct.} \bibinfo{year}{2015}), \bibinfo{pages}{32–36}.
\newblock
\showISSN{1072-5520}
\urldef\tempurl%
\url{https://doi.org/10.1145/2824892}
\showDOI{\tempurl}


\bibitem[H\"{o}\"{o}k et~al\mbox{.}(2015b)]%
        {Hook2015knowledge}
\bibfield{author}{\bibinfo{person}{Kristina H\"{o}\"{o}k}, \bibinfo{person}{Peter Dalsgaard}, \bibinfo{person}{Stuart Reeves}, \bibinfo{person}{Jeffrey Bardzell}, \bibinfo{person}{Jonas L\"{o}wgren}, \bibinfo{person}{Erik Stolterman}, {and} \bibinfo{person}{Yvonne Rogers}.} \bibinfo{year}{2015}\natexlab{b}.
\newblock \showarticletitle{Knowledge Production in Interaction Design} \emph{(\bibinfo{series}{CHI EA '15})}. \bibinfo{publisher}{Association for Computing Machinery}, \bibinfo{address}{New York, NY, USA}, \bibinfo{pages}{2429–2432}.
\newblock
\showISBNx{9781450331463}
\urldef\tempurl%
\url{https://doi.org/10.1145/2702613.2702653}
\showDOI{\tempurl}


\bibitem[H\"{o}\"{o}k and L\"{o}wgren(2012)]%
        {Hook2012}
\bibfield{author}{\bibinfo{person}{Kristina H\"{o}\"{o}k} {and} \bibinfo{person}{Jonas L\"{o}wgren}.} \bibinfo{year}{2012}\natexlab{}.
\newblock \showarticletitle{Strong Concepts: Intermediate-Level Knowledge in Interaction Design Research}.
\newblock \bibinfo{journal}{\emph{ACM Trans. Comput.-Hum. Interact.}} \bibinfo{volume}{19}, \bibinfo{number}{3}, Article \bibinfo{articleno}{23} (\bibinfo{date}{10} \bibinfo{year}{2012}), \bibinfo{numpages}{18}~pages.
\newblock
\showISSN{1073-0516}
\urldef\tempurl%
\url{https://doi.org/10.1145/2362364.2362371}
\showDOI{\tempurl}


\bibitem[Hsu et~al\mbox{.}(2018)]%
        {hsu2018botanical}
\bibfield{author}{\bibinfo{person}{Yuan-Yao Hsu}, \bibinfo{person}{Wenn-Chieh Tsai}, \bibinfo{person}{Wan-Chen Lee}, {and} \bibinfo{person}{Rung-Huei Liang}.} \bibinfo{year}{2018}\natexlab{}.
\newblock \showarticletitle{Botanical Printer: An Exploration on Interaction Design with Plantness}. In \bibinfo{booktitle}{\emph{Proceedings of the 2018 Designing Interactive Systems Conference}} (Hong Kong, China) \emph{(\bibinfo{series}{DIS '18})}. \bibinfo{publisher}{Association for Computing Machinery}, \bibinfo{address}{New York, NY, USA}, \bibinfo{pages}{1055–1068}.
\newblock
\showISBNx{9781450351980}
\urldef\tempurl%
\url{https://doi.org/10.1145/3196709.3196809}
\showDOI{\tempurl}


\bibitem[Jarvis et~al\mbox{.}(2012)]%
        {jarvis2012attention}
\bibfield{author}{\bibinfo{person}{Nadine Jarvis}, \bibinfo{person}{David Cameron}, {and} \bibinfo{person}{Andy Boucher}.} \bibinfo{year}{2012}\natexlab{}.
\newblock \showarticletitle{Attention to Detail: Annotations of a Design Process} \emph{(\bibinfo{series}{NordiCHI '12})}. \bibinfo{publisher}{Association for Computing Machinery}, \bibinfo{address}{New York, NY, USA}, \bibinfo{pages}{11–20}.
\newblock
\showISBNx{9781450314824}
\urldef\tempurl%
\url{https://doi.org/10.1145/2399016.2399019}
\showDOI{\tempurl}


\bibitem[Ju(2015)]%
        {ju2015design}
\bibfield{author}{\bibinfo{person}{Wendy Ju}.} \bibinfo{year}{2015}\natexlab{}.
\newblock \showarticletitle{The design of implicit interactions}.
\newblock \bibinfo{journal}{\emph{Synthesis Lectures on Human-Centered Informatics}} \bibinfo{volume}{8}, \bibinfo{number}{2} (\bibinfo{year}{2015}), \bibinfo{pages}{1--93}.
\newblock


\bibitem[Koeman et~al\mbox{.}(2014)]%
        {Koeman2014}
\bibfield{author}{\bibinfo{person}{Lisa Koeman}, \bibinfo{person}{Vaiva Kalnikait\.{e}}, \bibinfo{person}{Yvonne Rogers}, {and} \bibinfo{person}{Jon Bird}.} \bibinfo{year}{2014}\natexlab{}.
\newblock \showarticletitle{What Chalk and Tape Can Tell Us: Lessons Learnt for Next Generation Urban Displays}. In \bibinfo{booktitle}{\emph{Proceedings of The International Symposium on Pervasive Displays}} (Copenhagen, Denmark) \emph{(\bibinfo{series}{PerDis '14})}. \bibinfo{publisher}{ACM}, \bibinfo{address}{New York, NY, USA}, Article \bibinfo{articleno}{130}, \bibinfo{numpages}{6}~pages.
\newblock
\showISBNx{978-1-4503-2952-1}
\urldef\tempurl%
\url{https://doi.org/10.1145/2611009.2611018}
\showDOI{\tempurl}


\bibitem[Koskinen et~al\mbox{.}(2011)]%
        {Koskinen2011}
\bibfield{author}{\bibinfo{person}{Ilpo Koskinen}, \bibinfo{person}{John Zimmerman}, \bibinfo{person}{Thomas Binder}, \bibinfo{person}{Johan Redström}, {and} \bibinfo{person}{Stephan Wensveen}.} \bibinfo{year}{2011}\natexlab{}.
\newblock \bibinfo{booktitle}{\emph{Design Research Through Practice: From the Lab, Field, and Showroom}}. Vol.~\bibinfo{volume}{56}.
\newblock
\urldef\tempurl%
\url{https://doi.org/10.1109/TPC.2013.2274109}
\showDOI{\tempurl}


\bibitem[Lee et~al\mbox{.}(2019)]%
        {lee2019design}
\bibfield{author}{\bibinfo{person}{Wen-Ying Lee}, \bibinfo{person}{Yoyo Tsung-Yu Hou}, \bibinfo{person}{Cristina Zaga}, {and} \bibinfo{person}{Malte Jung}.} \bibinfo{year}{2019}\natexlab{}.
\newblock \showarticletitle{Design for serendipitous interaction: BubbleBot-bringing people together with bubbles}. In \bibinfo{booktitle}{\emph{2019 14th ACM/IEEE International Conference on Human-Robot Interaction (HRI)}}. IEEE, \bibinfo{pages}{759--760}.
\newblock


\bibitem[Lee and Jung(2020)]%
        {Lee2020}
\bibfield{author}{\bibinfo{person}{Wen-Ying Lee} {and} \bibinfo{person}{Malte Jung}.} \bibinfo{year}{2020}\natexlab{}.
\newblock \showarticletitle{Ludic-HRI: Designing Playful Experiences with Robots}. In \bibinfo{booktitle}{\emph{Companion of the 2020 ACM/IEEE International Conference on Human-Robot Interaction}} (Cambridge, United Kingdom) \emph{(\bibinfo{series}{HRI ’20})}. \bibinfo{publisher}{Association for Computing Machinery}, \bibinfo{address}{New York, NY, USA}, \bibinfo{pages}{582–584}.
\newblock
\showISBNx{9781450370578}
\urldef\tempurl%
\url{https://doi.org/10.1145/3371382.3377429}
\showDOI{\tempurl}


\bibitem[Lewis and Rieman(1993)]%
        {Lewis1993}
\bibfield{author}{\bibinfo{person}{Clayton Lewis} {and} \bibinfo{person}{John Rieman}.} \bibinfo{year}{1993}\natexlab{}.
\newblock \bibinfo{booktitle}{\emph{Task-Centered User Interface Design: A Practical Introduction}}.
\newblock \bibinfo{publisher}{University of Colorado, Boulder, Department of Computer Science}, \bibinfo{address}{Boulder, CO, USA}.
\newblock


\bibitem[Lockton et~al\mbox{.}(2020)]%
        {lockton2020sleep}
\bibfield{author}{\bibinfo{person}{Dan Lockton}, \bibinfo{person}{Tammar Zea-Wolfson}, \bibinfo{person}{Jackie Chou}, \bibinfo{person}{Yuhan~(Antonio) Song}, \bibinfo{person}{Erin Ryan}, {and} \bibinfo{person}{CJ Walsh}.} \bibinfo{year}{2020}\natexlab{}.
\newblock \bibinfo{booktitle}{\emph{Sleep Ecologies: Tools for Snoozy Autoethnography}}.
\newblock \bibinfo{publisher}{Association for Computing Machinery}, \bibinfo{address}{New York, NY, USA}, \bibinfo{pages}{1579–1591}.
\newblock
\showISBNx{9781450369749}
\urldef\tempurl%
\url{https://doi.org/10.1145/3357236.3395482}
\showURL{%
\tempurl}


\bibitem[L\"{o}wgren(2013)]%
        {Lowgren2013}
\bibfield{author}{\bibinfo{person}{Jonas L\"{o}wgren}.} \bibinfo{year}{2013}\natexlab{}.
\newblock \showarticletitle{Annotated Portfolios and Other Forms of Intermediate-Level Knowledge}.
\newblock \bibinfo{journal}{\emph{Interactions}} \bibinfo{volume}{20}, \bibinfo{number}{1} (\bibinfo{date}{Jan.} \bibinfo{year}{2013}), \bibinfo{pages}{30–34}.
\newblock
\showISSN{1072-5520}
\urldef\tempurl%
\url{https://doi.org/10.1145/2405716.2405725}
\showDOI{\tempurl}


\bibitem[Lupetti et~al\mbox{.}(2019)]%
        {Lupetti2019a}
\bibfield{author}{\bibinfo{person}{Maria~Luce Lupetti}, \bibinfo{person}{Roy Bendor}, {and} \bibinfo{person}{Elisa Giaccardi}.} \bibinfo{year}{2019}\natexlab{}.
\newblock \showarticletitle{Robot Citizenship: A Design Perspective}.
\newblock In \bibinfo{booktitle}{\emph{DeSForM19 Proceedings}}.
\newblock
\urldef\tempurl%
\url{https://doi.org/10.21428/5395bc37.595d1e58}
\showDOI{\tempurl}
\newblock
\shownote{https://desform19.pubpub.org/pub/robot-citizenship}.


\bibitem[Lupetti et~al\mbox{.}(2020)]%
        {workshop}
\bibfield{author}{\bibinfo{person}{Maria~Luce Lupetti}, \bibinfo{person}{Cristina Zaga}, {and} \bibinfo{person}{Nazli Cila}.} \bibinfo{year}{2020}\natexlab{}.
\newblock \bibinfo{title}{First international workshop on Designerly HRI Knowledge}.
\newblock
\newblock


\bibitem[Lupetti et~al\mbox{.}(2021)]%
        {Lupetti2021}
\bibfield{author}{\bibinfo{person}{Maria~Luce Lupetti}, \bibinfo{person}{Cristina Zaga}, {and} \bibinfo{person}{Nazli Cila}.} \bibinfo{year}{2021}\natexlab{}.
\newblock \showarticletitle{Designerly Ways of Knowing in HRI: Broadening the Scope of Design-Oriented HRI Through the Concept of Intermediate-Level Knowledge}. In \bibinfo{booktitle}{\emph{Proceedings of the 2021 ACM/IEEE International Conference on Human-Robot Interaction}} (Boulder, CO, USA) \emph{(\bibinfo{series}{HRI '21})}. \bibinfo{publisher}{Association for Computing Machinery}, \bibinfo{address}{New York, NY, USA}, \bibinfo{pages}{389–398}.
\newblock
\showISBNx{9781450382892}
\urldef\tempurl%
\url{https://doi.org/10.1145/3434073.3444668}
\showDOI{\tempurl}


\bibitem[Luria et~al\mbox{.}(2021)]%
        {LuriaHRI21}
\bibfield{author}{\bibinfo{person}{Michal Luria}, \bibinfo{person}{Marius Hoggenm\"{u}ller}, \bibinfo{person}{Wen-Ying Lee}, \bibinfo{person}{Luke Hespanhol}, \bibinfo{person}{Malte Jung}, {and} \bibinfo{person}{Jodi Forlizzi}.} \bibinfo{year}{2021}\natexlab{}.
\newblock \showarticletitle{Research through Design Approaches in Human-Robot Interaction}. In \bibinfo{booktitle}{\emph{Companion of the 2021 ACM/IEEE International Conference on Human-Robot Interaction}} (Boulder, CO, USA) \emph{(\bibinfo{series}{HRI '21 Companion})}. \bibinfo{publisher}{Association for Computing Machinery}, \bibinfo{address}{New York, NY, USA}, \bibinfo{pages}{685–687}.
\newblock
\showISBNx{9781450382908}
\urldef\tempurl%
\url{https://doi.org/10.1145/3434074.3444868}
\showDOI{\tempurl}


\bibitem[Luria et~al\mbox{.}(2019)]%
        {Luria2019}
\bibfield{author}{\bibinfo{person}{Michal Luria}, \bibinfo{person}{John Zimmerman}, {and} \bibinfo{person}{Jodi Forlizzi}.} \bibinfo{year}{2019}\natexlab{}.
\newblock \bibinfo{title}{Championing Research Through Design in HRI}.  (\bibinfo{date}{05} \bibinfo{year}{2019}).
\newblock
\newblock
\shownote{Paper presented at the CHI Conference on Human Factors in Computing Systems}.


\bibitem[Overgoor and Funk(2018)]%
        {Overgoor2018IdleBot}
\bibfield{author}{\bibinfo{person}{Caroline Overgoor} {and} \bibinfo{person}{Mathias Funk}.} \bibinfo{year}{2018}\natexlab{}.
\newblock \showarticletitle{IdleBot: Exploring the Design of Serendipitous Artifacts}. In \bibinfo{booktitle}{\emph{Proceedings of the 2018 ACM Conference Companion Publication on Designing Interactive Systems}} (Hong Kong, China) \emph{(\bibinfo{series}{DIS '18 Companion})}. \bibinfo{publisher}{Association for Computing Machinery}, \bibinfo{address}{New York, NY, USA}, \bibinfo{pages}{105–110}.
\newblock
\showISBNx{9781450356312}
\urldef\tempurl%
\url{https://doi.org/10.1145/3197391.3205420}
\showDOI{\tempurl}


\bibitem[Pataranutaporn et~al\mbox{.}(2020)]%
        {spaceworkshop}
\bibfield{author}{\bibinfo{person}{Pat Pataranutaporn}, \bibinfo{person}{Valentina Sumini}, \bibinfo{person}{Ariel Ekblaw}, \bibinfo{person}{Melodie Yashar}, \bibinfo{person}{Sandra Häuplik-Meusburger}, \bibinfo{person}{Susanna Testa}, \bibinfo{person}{Marianna Obrist}, \bibinfo{person}{Dorit Donoviel}, \bibinfo{person}{Joseph Paradiso}, {and} \bibinfo{person}{Pattie Maes}.} \bibinfo{year}{2020}\natexlab{}.
\newblock \bibinfo{title}{SpaceCHI : Human-Computer Interaction for Space Exploration}.
\newblock
\newblock


\bibitem[Rasmussen et~al\mbox{.}(2019)]%
        {rasmussen2019co}
\bibfield{author}{\bibinfo{person}{S\o{}ren Rasmussen}, \bibinfo{person}{Jeanette Falk~Olesen}, {and} \bibinfo{person}{Kim Halskov}.} \bibinfo{year}{2019}\natexlab{}.
\newblock \showarticletitle{Co-Notate: Exploring Real-Time Annotations to Capture Situational Design Knowledge}. In \bibinfo{booktitle}{\emph{Proceedings of the 2019 on Designing Interactive Systems Conference}} (San Diego, CA, USA) \emph{(\bibinfo{series}{DIS '19})}. \bibinfo{publisher}{Association for Computing Machinery}, \bibinfo{address}{New York, NY, USA}, \bibinfo{pages}{161–172}.
\newblock
\showISBNx{9781450358507}
\urldef\tempurl%
\url{https://doi.org/10.1145/3322276.3322310}
\showDOI{\tempurl}


\bibitem[Sengers and Gaver(2006)]%
        {Sengers}
\bibfield{author}{\bibinfo{person}{Phoebe Sengers} {and} \bibinfo{person}{Bill Gaver}.} \bibinfo{year}{2006}\natexlab{}.
\newblock \showarticletitle{Staying Open to Interpretation: Engaging Multiple Meanings in Design and Evaluation}. In \bibinfo{booktitle}{\emph{Proceedings of the 6th Conference on Designing Interactive Systems}} (University Park, PA, USA) \emph{(\bibinfo{series}{DIS '06})}. \bibinfo{publisher}{Association for Computing Machinery}, \bibinfo{address}{New York, NY, USA}, \bibinfo{pages}{99–108}.
\newblock
\showISBNx{1595933670}
\urldef\tempurl%
\url{https://doi.org/10.1145/1142405.1142422}
\showDOI{\tempurl}


\bibitem[Tran~Luciani et~al\mbox{.}(2018)]%
        {Luciani2018MachineLA}
\bibfield{author}{\bibinfo{person}{Danwei Tran~Luciani}, \bibinfo{person}{Martin Lindvall}, {and} \bibinfo{person}{Jonas L{\"o}wgren}.} \bibinfo{year}{2018}\natexlab{}.
\newblock \showarticletitle{Machine learning as a design material : a curated collection of exemplars for visual interaction}. In \bibinfo{booktitle}{\emph{DS 91: Proceedings of NordDesign 2018, Linköping, Sweden, 14th - 17th August 2018 :}} \emph{(\bibinfo{series}{NordDESIGN}, \bibinfo{number}{2018})}. \bibinfo{pages}{1--10}.
\newblock
\showISBNx{9789176851852}


\bibitem[Tuomi et~al\mbox{.}(2021)]%
        {tuomi2021spicing}
\bibfield{author}{\bibinfo{person}{Aarni Tuomi}, \bibinfo{person}{Iis~P Tussyadiah}, {and} \bibinfo{person}{Paul Hanna}.} \bibinfo{year}{2021}\natexlab{}.
\newblock \showarticletitle{Spicing up hospitality service encounters: the case of Pepper™}.
\newblock \bibinfo{journal}{\emph{International Journal of Contemporary Hospitality Management}} (\bibinfo{year}{2021}).
\newblock
\urldef\tempurl%
\url{https://doi.org/10.1108/IJCHM-07-2020-0739}
\showDOI{\tempurl}


\bibitem[Zamfirescu-Pereira et~al\mbox{.}(2021)]%
        {Zamfirescu2021}
\bibfield{author}{\bibinfo{person}{J.D. Zamfirescu-Pereira}, \bibinfo{person}{David Sirkin}, \bibinfo{person}{David Goedicke}, \bibinfo{person}{Ray LC}, \bibinfo{person}{Natalie Friedman}, \bibinfo{person}{Ilan Mandel}, \bibinfo{person}{Nikolas Martelaro}, {and} \bibinfo{person}{Wendy Ju}.} \bibinfo{year}{2021}\natexlab{}.
\newblock \showarticletitle{Fake It to Make It: Exploratory Prototyping in HRI}. In \bibinfo{booktitle}{\emph{Companion of the 2021 ACM/IEEE International Conference on Human-Robot Interaction}} (Boulder, CO, USA) \emph{(\bibinfo{series}{HRI '21 Companion})}. \bibinfo{publisher}{Association for Computing Machinery}, \bibinfo{address}{New York, NY, USA}, \bibinfo{pages}{19–28}.
\newblock
\showISBNx{9781450382908}
\urldef\tempurl%
\url{https://doi.org/10.1145/3434074.3446909}
\showDOI{\tempurl}


\bibitem[Zimmerman and Forlizzi(2008)]%
        {Zimmerman2008}
\bibfield{author}{\bibinfo{person}{John Zimmerman} {and} \bibinfo{person}{Jodi Forlizzi}.} \bibinfo{year}{2008}\natexlab{}.
\newblock \showarticletitle{The Role of Design Artifacts in Design Theory Construction}.
\newblock \bibinfo{journal}{\emph{Artifact}} \bibinfo{volume}{2}, \bibinfo{number}{1} (\bibinfo{year}{2008}), \bibinfo{pages}{41--45}.
\newblock
\urldef\tempurl%
\url{https://doi.org/10.1080/17493460802276893}
\showDOI{\tempurl}
\showeprint{https://doi.org/10.1080/17493460802276893}


\bibitem[Zimmerman et~al\mbox{.}(2007)]%
        {zimmerman2007research}
\bibfield{author}{\bibinfo{person}{John Zimmerman}, \bibinfo{person}{Jodi Forlizzi}, {and} \bibinfo{person}{Shelley Evenson}.} \bibinfo{year}{2007}\natexlab{}.
\newblock \showarticletitle{Research through design as a method for interaction design research in HCI}. In \bibinfo{booktitle}{\emph{Proceedings of the SIGCHI conference on Human factors in computing systems}}. \bibinfo{pages}{493--502}.
\newblock


\bibitem[Zimmerman et~al\mbox{.}(2010)]%
        {Zimmerman2010}
\bibfield{author}{\bibinfo{person}{John Zimmerman}, \bibinfo{person}{Erik Stolterman}, {and} \bibinfo{person}{Jodi Forlizzi}.} \bibinfo{year}{2010}\natexlab{}.
\newblock \showarticletitle{An Analysis and Critique of Research through Design: Towards a Formalization of a Research Approach}. In \bibinfo{booktitle}{\emph{Proceedings of the 8th ACM Conference on Designing Interactive Systems}} (Aarhus, Denmark) \emph{(\bibinfo{series}{DIS '10})}. \bibinfo{publisher}{Association for Computing Machinery}, \bibinfo{address}{New York, NY, USA}, \bibinfo{pages}{310–319}.
\newblock
\showISBNx{9781450301039}
\urldef\tempurl%
\url{https://doi.org/10.1145/1858171.1858228}
\showDOI{\tempurl}


\end{thebibliography}

\end{document}